\documentclass[12pt,preprint]{aastex}
\shorttitle{H$_2$ in Unstable Disks}
\shortauthors{Boley et al.}

\begin{document}
\title{The Internal Energy for Molecular Hydrogen in Gravitationally Unstable Protoplanetary Disks}

\altaffiltext{1}{Indiana University, Astronomy Department, 727 E.\ Third St., Bloomington, IN 47405-7105; acboley@astro.indiana.edu}
\altaffiltext{2}{University of Leeds, School of Physics and Astronomy, Leeds LS2 9JT, UK}

\author{Aaron C.\ Boley\altaffilmark{1}, Thomas W.~Hartquist\altaffilmark{2}, Richard H.~Durisen\altaffilmark{1}, Scott Michael\altaffilmark{1}}




\begin{abstract}
The gas equation of state may be one of the critical factors for the disk instability theory of gas giant planet formation. This letter addresses the treatment of H$_2$ in hydrodynamical simulations of gravitationally unstable disks.  In our discussion, we point out possible consequences of erroneous specific internal energy relations, approximate specific internal energy relations with discontinuities, and assumptions of constant $\Gamma_1$.  In addition, we consider whether the ortho/para ratio for H$_2$ in protoplanetary disks should be treated dynamically as if the species are in equilibrium. Preliminary simulations indicate that the correct treatment is particularly critical for the study of gravitational instability when $T = 30$-$50$ K.
\end{abstract}


\section{INTRODUCTION\label{intro}}

Researchers disagree on several key issues of disk evolution, particularly regarding fragmentation and planet formation, despite numerous studies of gravitational instabilities in protoplanetary disks (see Durisen et al.~2006 for a review). Recent three-dimensional radiative hydrodynamics simulations of protoplanetary disks with gravitational instabilities (GIs) demonstrate disparate evolutions, and the differences involve the importance of convection in disks, the dependence of disk cooling on metallicity, and the stability of disks to fragmentation and clump formation \citep{boss2005,cai_letter_2006,paperIII,mayer2006letter}.

These disparities may be due to differences in the treatments of radiative transfer and the assumed equation of state.  Our own group has undertaken systematic improvements to the Indiana University Hydrodynamics Group code in these two areas. Boley et al.~(in preparation) will present a new scheme for 3D radiative cooling in disks that passes a series of accuracy and reliability tests, which are suggested as a new standard for disk stability studies.  Because the equation of state can severely change the outcome of a simulation \citep{pickett1998,pickett2000}, we have also implemented an internal energy that takes into account the translational, rotational, and vibrational states of H$_2$. During its development, we noticed that in all treatments to date of planet formation by disk instability, the effects of the rotational states of H$_2$ have been, at best, only poorly approximated.  The purpose of this Letter is to draw attention to possible consequences of various approximations for the internal energy of H$_2$ that are in the literature and to disseminate the correct treatment for H$_2$.  We have begun a series of simulations to investigate the severity of using the incorrect internal energies. We mention only preliminary results in this Letter, but the simulations will be fully discussed in a future paper.  

This Letter is laid out as follows: In \S 2 we outline the correct treatment of H$_2$ under simple conditions, i.e., no dissociation and no ionization.  We discuss the ortho/para hydrogen ratio for optically thick protoplanetary disks in \S 3, and we summarize the main points of the letter in \S 4.

\section{THERMODYNAMICS}

In this section we summarize relevant thermodynamic relations.  For an in-depth discussion, we refer the reader to \citet{pathria}.  Consider the following thermodynamic properties of an ideal gas: Let $E$ be the internal energy for $N$ particles, $e$ the specific internal energy, $\epsilon$ the internal energy density, $p$ the pressure, $T$ the gas temperature, $\rho$ the gas density, $\mu$ the mean molecular weight in proton masses, $c_v$ the specific heat capacity at constant volume, $Z$ the partition function for the ensemble, $z$ the partition function for a single particle, and $R = k/m_p$, where $k$ is Boltzmann's constant and $m_p$ is the proton mass.  We only consider independent contributions to the partition function from translation, rotation, and vibration represented by
$ Z = Z_{\rm tran}Z_{\rm rot}Z_{\rm vib}= z^N=
\left( z_{\rm tran} z_{\rm rot} z_{\rm vib}\right)^N$.
The internal energy $E$ and the specific internal energy $e$ can be calculated by
\begin{equation}E=NkT^2\frac{\partial \ln z}{\partial T};~e=\frac{R}{\mu}T^2\frac{\partial \ln z}{\partial T}\label{eq1}\end{equation}
for constant $\rho$. Because the gas is ideal, $c_v = d e/d T$.
If $c_v$ is constant, then $e = c_v T.$
\citet{black_bodenheimer1975} calculate $c_v$ from the Helmholtz free energy, which is valid, but then assume $e=c_v T$, which is invalid, because $c_v$ is dependent on $T$.  Other authors have followed suit \citep[e.g.,][]{whitehouse2006}, and this assumption could be troublesome for gas dynamics in a hydrodynamics simulation (see below).  We do note that the severity of this error may depend on the state variables evolved in a given code, and only the authors who employ $e=c_v T$ will be able to say in detail how it affects their simulations.

\subsection{Molecular Hydrogen}

Molecular hydrogen exists as para hydrogen and as orthohydrogen where the proton spins are antiparallel and parallel, respectively.  The partition function for parahydrogen is
\begin{equation}z_p=\sum_{j_{\rm even}}\left(2j + 1\right) \exp\left( -j\left(j+1\right)\theta_{\rm rot}/T\right),
\label{eq2}\end{equation}
and the partition function for orthohydrogen is
\begin{equation}z_o=\sum_{j_{\rm odd}}3\left(2j + 1\right) \exp\left( -j\left(j+1\right)\theta_{\rm rot}/T\right)
\label{eq3},\end{equation}
where $\theta_{\rm rot}=85.4$ K \citep{black_bodenheimer1975}.  When the two species are in equilibrium, $z_{\rm rot} = z_p + z_o$.   However, the ortho/para ratio (b:a) could also be frozen if no efficient mechanism for converting between the species is available. This leads to
$z_{\rm rot} = z_p^{\left(a/\left(a+b\right)\right)} z_o'^{\left(b/\left( a + b\right)\right)}$,
where
$z_o' =z_o \exp\left( 2\theta_{\rm rot}/T \right)$.
The additional exponential is required in the orthohydrogen partition function when the ortho and para species are at some fixed ratio to ensure that rotation only contributes to the internal energy once the rotational states are excited, i.e., $z_o'\rightarrow 1$ as $T \rightarrow 0$.

To consider the vibrational states, we approximate the molecule as an infinitely deep harmonic oscillator, where
\begin{equation} z_{\rm vib} = \frac{1}{1-\exp\left(-\theta_{\rm vib}/T\right)};\label{eq4}\end{equation}
$\theta_{\rm vib} = 5987$ K \citep{draineetal1983}.  Because we are only interested in temperatures $T\lesssim1500$ K where dissociation of H$_2$ is insignificant, we can ignore differences between equation (4) and a proper $z_{\rm vib}$, which would take into account the anharmonicity of the molecule and that the molecule has a finite number of vibrationally excited states. 

We can use equation (1) to write the specific internal energy for H$_2$
\begin{equation} e\left({\rm H}_2\right) = \frac{R}{2}\left( \frac{3}{2}T + \frac{T^2}{z_{\rm rot}}\frac{\partial z_{\rm rot}}{\partial T} + \theta_{\rm vib}\frac{\exp\left( -\theta_{\rm vib}/T\right)}{1-\exp\left( -\theta_{\rm vib}/T\right)}\right).\label{5}
\end{equation}

When the gas is ideal and dissociation and ionization can be ignored, the first adiabatic exponent $\Gamma_1 = 1+ R/\mu c_v = c_p/c_v = \gamma$ \citep{cox_giuli1968}.  Figure 1 indicates the $\Gamma_1$ profiles for the equilibrium case, pure parahydrogen, and a 3:1 mix, which is consistent with Figure 2 of \citet{decamplietal1978}, as it should be because our derivation of $c_v$ is equivalent to theirs. The specific internal energy profiles as calculated by equation (5) are shown in Figure 2 by solid lines.  Curves for  $c_v T$ are also shown in Figure 2 by dashed lines.  The approximation $e=c_v T$ gives quite different behavior from the correct $e$, e.g., the incorrect 3:1 curve most closely follows the correct pure parahydrogen curve.  The offset of the correct 3:1 mix profile is due to the energy stored in the parallel spins of the protons.  
The dotted line represents the curve $e=3/4 R T$ for $T < 100$ K and $e=5/4 R T$ for $T\ge100$ K, which is the energy equation used by \citet{boss1984,boss2001,boss2002,boss2005}.  Although this may be a reasonable approximation for $e$ when the hydrogen species are in equilibrium, this approximation could be troublesome for the hydrodynamics.  This is because the discontinuity guarantees that $c_v$ becomes very large near 100 K, which forces $\Gamma_1$ to unity.  Such discontinuities may artificially favor gas giant formation by disk instability, because isothermal ($\Gamma_1$) gas behavior favors fragmentation \citep{boss1997,boss2000,pickett1998}.  Indeed, the clumping that is reported by \citet{boss2001} occurs at $T\approx100$ K (see Boss's Figure 2).  Finally, a constant $\Gamma_1$ approximation \citep{pickett2003,lodato_rice2004,mayer2004,rice2003,mnras364l56,mayer2006letter,mejia2005,cai_letter_2006,paperIII,mayer2006letter} poorly represents $e$ in the temperature regime where Jupiter probably formed; neither $\Gamma_1=5/3$ or 7/5 can be assumed confidently.

The dynamical effects that could result from assuming $e=c_v T$ are evaluated by taking the temperature derivative of the dashed curves in Figure 2.  The results are displayed in Figure 3, and the curves depart drastically from the correct $\Gamma_1$ profiles.   As mentioned above, $c_v$ can be calculated from the Helmholtz free energy as is done by \citet{black_bodenheimer1975}.  This makes it possible to compute $\Gamma_1$ from $c_v$ correctly but then evolve the gas with an erroneous effective specific heat because many hydrodynamics codes evolve $\epsilon=\rho e$ \citep[e.g.,][]{black_bodenheimer1975,boss1984,boss2001,monaghan1992,stone_norman1992,pickett1995,wadsleyetal2004}.  If the ideal gas law $p=R/\mu\rho T$ is assumed as well, then effective $\Gamma_1$ profiles like those shown in Figure 3 seem to be unavoidable when $e=c_v T$ is assumed for a temperature dependent $c_v$.  Because fragmentation becomes more likely as $\Gamma_1$ becomes smaller (Rice et al.~2005; Michael et al.~in preparation), the $e=c_v T$ assumption should artificially make fragmentation more likely in some temperature regimes and less likely in others.

  Preliminary simulations of a disk with solar composition indicate that when GIs activate between 30 and 50K for an equilibrium ortho-para mixture, the $e=c_v T$ simulation 
evolves more rapidly and has a more flocculent spiral structure than the correct $e$ simulation
for the same cooling rates.  In addition, denser substructures form in some spiral arms of the $e=c_v T$ simulation throughout the simulation, while dense substructures only form during the burst of the correct $e$ simulation\footnote{The evolution of these simulations can be viewed at http://westworld.astro.indiana.edu/.  Click on the link titled ``H2 ortho-para equilibrium tests'' under the ``Movies'' tab.}. When the instabilities occur outside this temperature regime, the differences are diminished. 

\section{THE MOLECULAR HYDROGEN ORTHO/PARA RATIO}

As indicated in \S 2, the dynamical behavior of the gas is dependent on the ortho/para ratio and whether the species are in equilibrium for all $T$.  This ratio for various astrophysical conditions has been addressed by several authors \citep[e.g.,][]{osterbrock1962,dalgarnoetal1973,decamplietal1978,flower_watt1984,sternberg_neufeld1999,fuente1999,rodriguez-fernandez2000,floweretal2006},  typically in the context of interstellar clouds or photodissociation regions.  However, for plausible Solar Nebula conditions, the ortho/para ratio has been inadequately addressed; for example, \citet{decamplietal1978} used an estimate for the H$^+$ number density that was derived originally to give the total gas phase ion number density in gas for which dissociative recombination dominates the removal of ions.  At protoplanetary disk number densities, however, ion removal should be primarily on grain surfaces if the ratio of grain surface area to hydrogen nucleon number density is the same as it is in diffuse interstellar clouds.

For protoplanetary disk conditions, the conversion between ortho and parahydrogen is principally due to  protonated ions such as H$^+_3$.
 Consequently, we will assume that all ionizations lead to H$^+_3$ formation.
Another possible conversion mechanism is through interactions between H$_2$ and grains; however, this conversion might only be significant when the temperature drops below about 30 K \citep{lebourlot2000}.

Consider the balance between $\rm H_3^+$ production by cosmic rays (CR) and H$_3^+$ depletion by dust grains:
\begin{equation} \zeta n\left({\rm H}_2\right)= n_g\pi a^2 v n_i, \end{equation}  
 where $n\left({\rm H_2}\right)$, $n_i$, and $n_g$ are the $\rm H_2$, $\rm H_3^+$, and grain number densities, respectively, $\zeta$ is the ionization rate by CRs and other energetic particles (EP), $a$ is the average radius of the grains, and $v$ is the thermal velocity of $\rm H_3^+$.  If we assume standard interstellar extinction, then $\sigma=n\left({\rm H_2}\right)/n_g \pi a^2 \approx 10^{21}$ cm$^{-2}$, but as we discuss below, this number is ambiguous.  The $\zeta$ appropriate for a protoplanetary disk is also ambiguous.  Cosmic rays and stellar EPs are important in ionizing the disk surface \citep{desch2004,dullemond_ppv}, but because these particles are attenuated exponentially with a scale length of about 100 g cm$^{-2}$ \citep{umebayashi_nakano1981}, stellar EPs probably do not contribute to $n_i$.  Moreover, protostellar winds could lead to a significant reduction of $\zeta$ in analogy to CR modulation by the solar wind \citep{webber1998}. However, at a surface density of roughly 380 g cm$^{-2}$, EP production by $^{26}$Al decay is as important as CRs with $\zeta\sim10^{-19}~\rm s^{-1}$ \citep{stepinski1992}. It is likely that $10^{-19}~{\rm s}^{-1}< \zeta<10^{-17} {\rm\ s}^{-1}$.   For our estimate we adopt the interstellar rate $\zeta = 10^{-17} {\rm\ s}^{-1}$ \citep{spitzer_tomasko1968}.  Using these numbers in equation (6) and adopting a thermal velocity of 1 km s$^{-1}$, $n_i\approx 0.1\rm\ cm^{-3}$.  By adopting a collisional rate coefficient $\alpha=1\times10^{-9}\rm\ cm^3\ s^{-1}$ for the H$_3^+$ interaction with H$_2$ \citep{walmsleyetal2004}, the lower limit timescale for ortho and para hydrogen to reach equilibrium is
$t_e=\left( \alpha n_i\right)^{-1} = 300$ yr. 

The equilibrium timescale is short enough that the ortho/para ratio can thermalize in the lifetime of a disk, but the equilibrium timescale is longer than the dynamical timescale inside about 40 AU: ortho and parahydrogen should be treated as independent species for hydrodynamical simulations of young protoplanetary disks. 

What ortho/para ratio should a dynamicist assume for gravitationally unstable protoplanetary disk simulations? The answer is uncertain. Vertical and radial stirring induced by shock bores \citep{boleyshockbores}, which could possibly lead to mixing of the low altitude disk interior with the high-altitude photodissociation region in the disk atmosphere \citep{dullemond_ppv}, will transport gas through different temperature regimes on dynamic timescales.  This could lead to nonthermalized ortho/para ratios like those that are measured from H$_2$ rotational transition lines in some photodissociation regions \citep{fuente1999,rodriguez-fernandez2000} and in Neptune's stratosphere \citep{fouchetetal2003}.  Moreover, accretion of the outer disk will bring material with a cold history into warmer regions of the disk.  It is unclear whether the ortho/para ratio of, say, 15 K gas will be thermalized with $z_o/z_p\approx 0$ or whether the ortho/para ratio will be 3:1, which is the expected ratio for H$_2$ formation on cold grains \citep{floweretal2006}. 
Unfortunately, the ortho/para ratio may be critical to the evolution of a protoplanetary disk.  As can be seen in Figure 1, the pure parahydrogen mix has a $\Gamma_1$ that approaches 4/3 for $T\approx160$ K. 
This could make the 160 K regime the most likely region of the disk to fragment because, as $\Gamma_1$ decreases, it becomes harder for the gas to support itself against local gravitational and hydrodynamic stresses (Rice et al.~2005; Michael et al.~in preparation).   Hydrodynamicists need to consider ortho/para ratios between pure parahydrogen and 3:1 because of our ignorance of this ratio in protoplanetary disks. 

 The above discussion is based on the assumption that $\sigma\approx 10^{21}~\rm cm^{-2}$, which is probably reasonable for very young protoplanetary disks but may not be reasonable for disks with ages of about 1 Ma.  Grain growth and dust settling may significantly lower the value of $\sigma$ by depleting the total grain area \citep[e.g.,][]{sanoetal2000}.  Because models of T Tauri disks must take into account the effects of grain growth in order to match observed spectral energy distributions \citep{dalessio2001,dalessio2006,furlanetal2006}, there may be a period in a disk's evolution when the ortho and parahydrogen change from dynamically independent species to species in statistical equilibrium.  Such a transition may also take place at certain radii in a disk, e.g., near edges of a dead zone \citep{gammie1996}.  As indicated by Figure 1, a transition to statistical equilibrium could have significant dynamical consequences for disk evolution and may induce clump formation by GIs.

\section{SUMMARY}
The effects of the rotational states of H$_2$ must be explicitly modeled in hydrodynamical simulations of protoplanetary disks.  Constant $\Gamma_1$ approximations are insufficient for modeling the dynamic behavior of the gas because they ignore the transition from a $\Gamma_1=5/3$ gas to a $\Gamma_1=7/5$ gas. This transition probably took place near Jupiter's location in the young Solar Nebula.  Discontinuous $e(T)$ assumptions or the $e=c_v T$ approximation could lead to severe errors in the behavior of the gas, including artificially driving $\Gamma_1\rightarrow 1$. 

Our estimates indicate that ortho and parahydrogen should be treated as separate species in hydrodynamics simulations of young protoplanetary disks because the timescale to thermalize the ortho/para ratio is longer than a dynamic timescale.  However, a reasonable approximation to the ortho/para ratio in young protoplanetary disks is ambiguous.  Moreover, as a disk evolves, there may be a transition in the behavior of ortho and parahydrogen from independent species to statistical equilibrium, which might trigger disk fragmentation.  Hydrodynamicists should consider a range of ortho/para ratios and the equilibrium case until this ambiguity is resolved.

\acknowledgements{We thank J.M.C.~Rawlings, A.~Dalgarno, and an anonymous referee for useful discussions and comments.  A.C.B.\ was supported by a NASA Graduate Student Researchers Program fellowship. R.H.D.'s and S.M.'s contributions were supported by NASA grant NNG05GN11G. The Indiana-Leeds collaboration has been supported in part by PPARC grants funding travel to and subsistence costs in Leeds.  }

\clearpage

\begin{figure}
\begin{center}
\includegraphics[width=10cm]{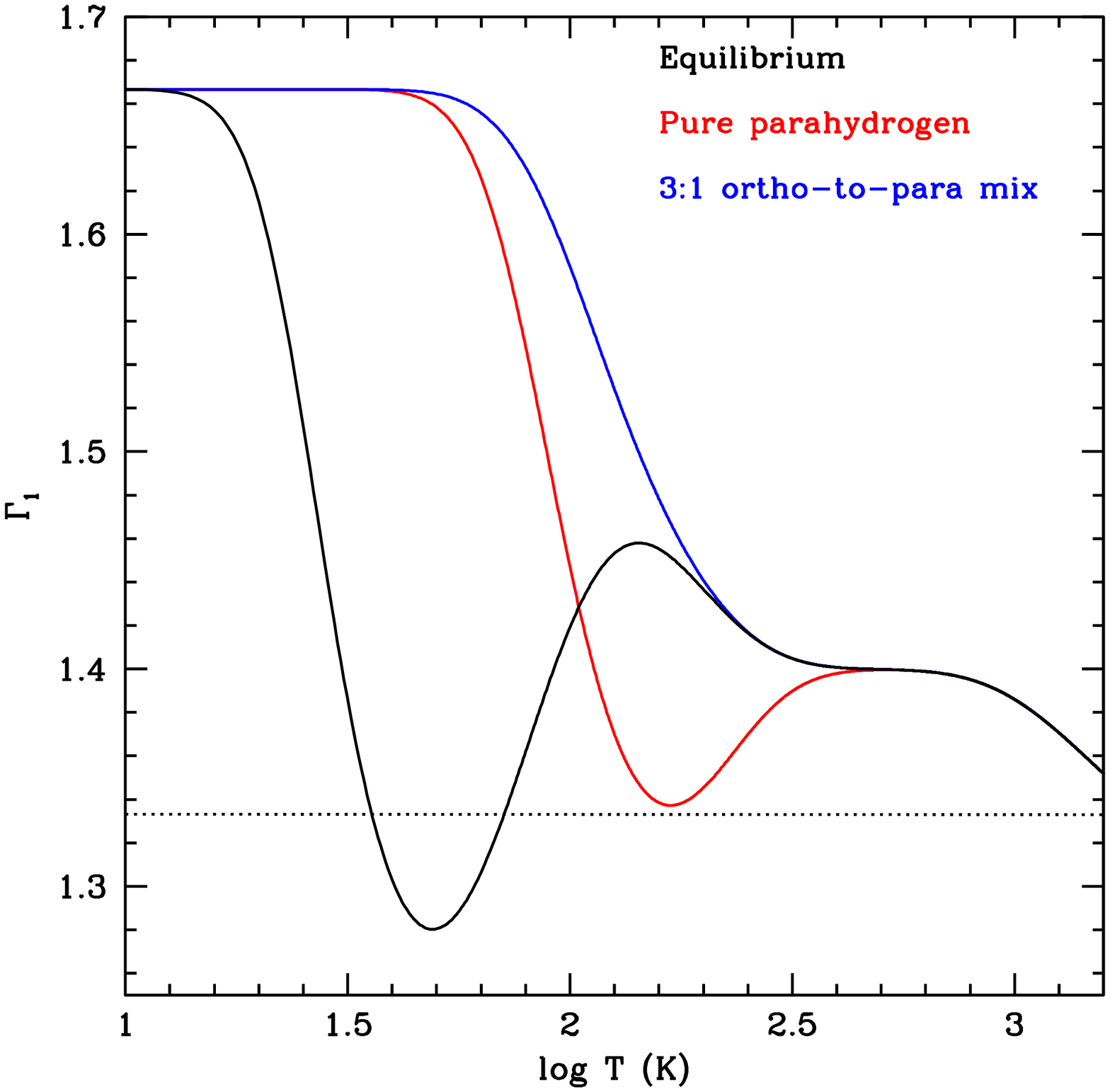}
\caption{Profiles of $\Gamma_1$ for an equilibrium mix (black), parahydrogen (red), and a 3:1 ortho/para ratio mix (blue).  The dotted line indicates $\Gamma_1=4/3$; this figure is similar to Figure 2 of \citep{decamplietal1978}. [Profiles of $\Gamma_1$ for an equilibrium mix (solid), parahydrogen (short dash), and a 3:1 ortho/para ratio mix (long dash).  The dotted line indicates $\Gamma_1=4/3$; this figure is similar to Figure 2 of \citep{decamplietal1978}.]}
\label{f1}
\end{center}
\end{figure}

\clearpage

\begin{figure}
\begin{center}
\includegraphics[width=10cm]{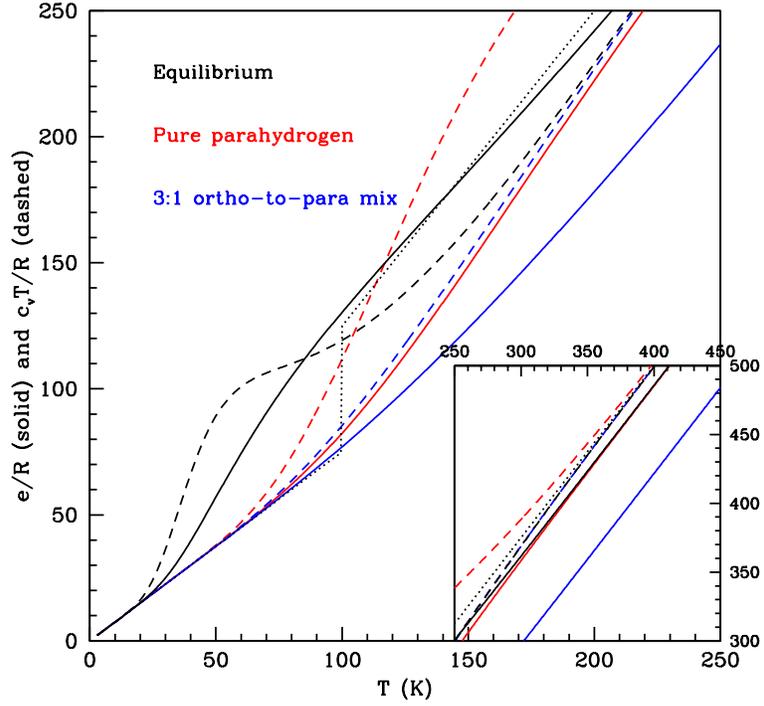}
\caption{Specific internal energy profiles.  The solid lines indicate the $e$ profiles as calculated by equation (5), while the dashed lines indicate $c_v T$.  The dotted line indicates a discontinuous $e$ profile where $e=3/4RT$ for $T<100$ K and $e=5/4RT$ for $T \ge 100$ K, as used by \citet{boss1984}.  The inset shows the behavior of the profiles at higher temperatures with the same units for the ordinate and abscissa. }
\label{f2}
\end{center}
\end{figure}

\clearpage

\begin{figure}
\begin{center}
\includegraphics[width=10cm]{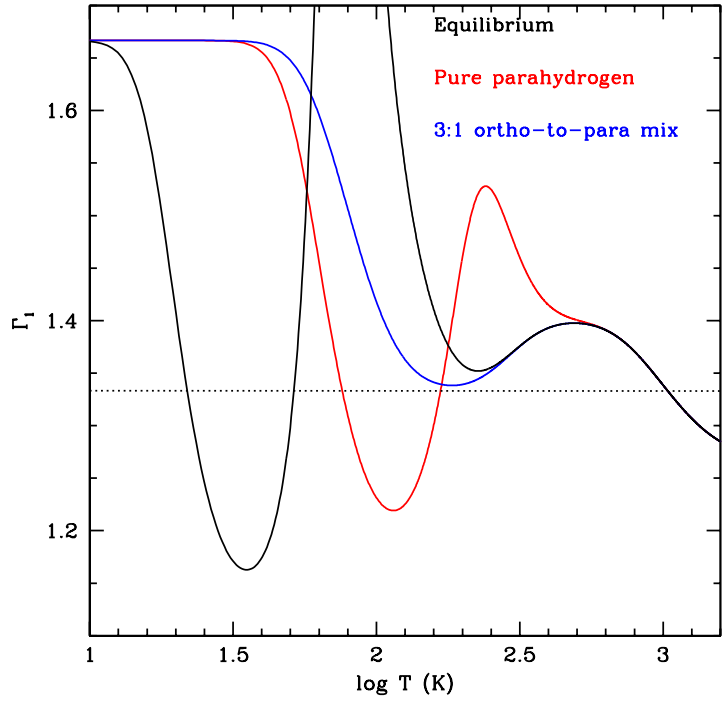}
\caption{Profiles of $\Gamma_1$ calculated by taking the temperature derivative of the dashed curves in Figure 2.  These profiles vary largely from the correct profiles in Figure 1.  The equilibrium curve's maximum is at about $\Gamma_1=2.3$.  }
\label{f3}
\end{center}
\end{figure}


\clearpage

\begin{thebibliography}{59}
\expandafter\ifx\csname natexlab\endcsname\relax\def\natexlab#1{#1}\fi


\bibitem[{{Black} \& {Bodenheimer}(1975)}]{black_bodenheimer1975}
{Black}, D.~C. \& {Bodenheimer}, P. 1975, \apj, 199, 619

\bibitem[{{Boley} \& {Durisen}(2006)}]{boleyshockbores}
{Boley}, A.~C. \& {Durisen}, R.~H. 2006, \apj, 641, 534

\bibitem[{{Boley} {et~al.}(2006){Boley}, {Mej{\'{\i}}a}, {Durisen}, {Cai},
  {Pickett}, \& {D'Alessio}}]{paperIII}
{Boley}, A.~C., {Mej{\'{\i}}a}, A.~C., {Durisen}, R.~H., {Cai}, K., {Pickett},
  M.~K., \& {D'Alessio}, P. 2006, ApJ, in press

\bibitem[{{Boss}(1984)}]{boss1984}
{Boss}, A.~P. 1984, \apj, 277, 768

\bibitem[{{Boss}(1997)}]{boss1997}
---. 1997, Science, 276, 1836

\bibitem[{{Boss}(1998)}]{boss1998}
---. 1998, ApJ, 503, 923

\bibitem[{{Boss}(2000)}]{boss2000}
---. 2000, \apjl, 536, L101

\bibitem[{{Boss}(2001)}]{boss2001}
---. 2001, ApJ, 563, 367

\bibitem[{{Boss}(2002)}]{boss2002}
---. 2002, ApJ, 576, 462

\bibitem[{{Boss}(2005)}]{boss2005}
---. 2005, ApJ, 629, 535

\bibitem[{{Cai} {et~al.}(2006){Cai}, {Durisen}, {Michael}, {Boley},
  {Mej{\'{\i}}a}, {Pickett}, \& {D'Alessio}}]{cai_letter_2006}
{Cai}, K., {Durisen}, R.~H., {Michael}, S., {Boley}, A.~C., {Mej{\'{\i}}a},
  A.~C., {Pickett}, M.~K., \& {D'Alessio}, P. 2006, ApJL, 636, L149

\bibitem[{{Cox} \& {Giuli}(1968)}]{cox_giuli1968}
{Cox}, J.~P. \& {Giuli}, R.~T. 1968, {Principles of stellar structure} (New
  York, Gordon and Breach [1968])

\bibitem[{{D'Alessio} {et~al.}(2001){D'Alessio}, {Calvet}, \&
  {Hartmann}}]{dalessio2001}
{D'Alessio}, P., {Calvet}, N., \& {Hartmann}, L. 2001, ApJ, 553, 321

\bibitem[{{D'Alessio} {et~al.}(2006){D'Alessio}, {Calvet}, {Hartmann},
  {Franco-Hern{\'a}ndez}, \& {Serv{\'{\i}}n}}]{dalessio2006}
{D'Alessio}, P., {Calvet}, N., {Hartmann}, L., {Franco-Hern{\'a}ndez}, R., \&
  {Serv{\'{\i}}n}, H. 2006, \apj, 638, 314

\bibitem[{{Dalgarno} {et~al.}(1973){Dalgarno}, {Black}, \&
  {Weisheit}}]{dalgarnoetal1973}
{Dalgarno}, A., {Black}, J.~H., \& {Weisheit}, J.~C. 1973, \aplett, 14, 77

\bibitem[{{Decampli} {et~al.}(1978){Decampli}, {Cameron}, {Bodenheimer}, \&
  {Black}}]{decamplietal1978}
{Decampli}, W.~M., {Cameron}, A.~G.~W., {Bodenheimer}, P., \& {Black}, D.~C.
  1978, \apj, 223, 854

\bibitem[{{Desch}(2004)}]{desch2004}
{Desch}, S.~J. 2004, \apj, 608, 509

\bibitem[{{Draine} {et~al.}(1983){Draine}, {Roberge}, \&
  {Dalgarno}}]{draineetal1983}
{Draine}, B.~T., {Roberge}, W.~G., \& {Dalgarno}, A. 1983, \apj, 264, 485

\bibitem[{{Dullemond} {et~al.}(2007){Dullemond}, {Hollenbach}, {Kamp}, \&
  {D'Alessio}}]{dullemond_ppv}
{Dullemond}, C.~P., {Hollenbach}, D., {Kamp}, I., \& {D'Alessio}, P.~D. 2007,
  in Protostars and Planets V

\bibitem[{{Durisen} {et~al.}(2007){Durisen}, {Boss}, {Mayer}, {Nelson},
  {Quinn}, \& {Rice}}]{durisen_ppv_chapter}
{Durisen}, R.~H., {Boss}, A.~P., {Mayer}, L., {Nelson}, A.~F., {Quinn}, T., \&
  {Rice}, W.~K.~M. 2007, in Protostars and Planets V

\bibitem[{{Flower} {et~al.}(2006){Flower}, {Pineau Des For{\^e}ts}, \&
  {Walmsley}}]{floweretal2006}
{Flower}, D.~R., {Pineau Des For{\^e}ts}, G., \& {Walmsley}, C.~M. 2006, \aap,
  449, 621

\bibitem[{{Flower} \& {Watt}(1984)}]{flower_watt1984}
{Flower}, D.~R. \& {Watt}, G.~D. 1984, \mnras, 209, 25

\bibitem[{{Fouchet} {et~al.}(2003){Fouchet}, {Lellouch}, \&
  {Feuchtgruber}}]{fouchetetal2003}
{Fouchet}, T., {Lellouch}, E., \& {Feuchtgruber}, H. 2003, Icarus, 161, 127

\bibitem[{{Fuente} {et~al.}(1999){Fuente}, {Mart{\'{\i}}n-Pintado},
  {Rodr{\'{\i}}guez-Fern{\'a}ndez}, {Rodr{\'{\i}}guez-Franco}, {de Vicente}, \&
  {Kunze}}]{fuente1999}
{Fuente}, A., {Mart{\'{\i}}n-Pintado}, J., {Rodr{\'{\i}}guez-Fern{\'a}ndez},
  N.~J., {Rodr{\'{\i}}guez-Franco}, A., {de Vicente}, P., \& {Kunze}, D. 1999,
  \apjl, 518, L45

\bibitem[{{Furlan} {et~al.}(2006){Furlan}, {Hartmann}, {Calvet}, {D'Alessio},
  {Franco-Hern{\'a}ndez}, {Forrest}, {Watson}, {Uchida}, {Sargent}, {Green},
  {Keller}, \& {Herter}}]{furlanetal2006}
{Furlan}, E., {Hartmann}, L., {Calvet}, N., {D'Alessio}, P.,
  {Franco-Hern{\'a}ndez}, R., {Forrest}, W.~J., {Watson}, D.~M., {Uchida},
  K.~I., {Sargent}, B., {Green}, J.~D., {Keller}, L.~D., \& {Herter}, T.~L.
  2006, \apjs, 165, 568

\bibitem[{{Gammie}(1996)}]{gammie1996}
{Gammie}, C.~F. 1996, ApJ, 457, 355

\bibitem[{{Gammie}(2001)}]{gammie2001}
---. 2001, ApJ, 553, 174

\bibitem[{{Le Bourlot}(2000)}]{lebourlot2000}
{Le Bourlot}, J. 2000, \aap, 360, 656
 
\bibitem[{{Lodato} \& {Rice}(2004)}]{lodato_rice2004}
{Lodato}, G. \& {Rice}, W.~K.~M. 2004, MNRAS, 351, 630

\bibitem[{{Mayer} {et~al.}(2006){Mayer}, {Graeme}, {Thomas}, \&
  {Wadsley}}]{mayer2006letter}
{Mayer}, L., {Lufkin}, G., {Quinn}, T., \& {Wadsley}, J. 2006,
  astro-ph/0606361

\bibitem[{{Mayer} {et~al.}(2004){Mayer}, {Quinn}, {Wadsley}, \&
  {Stadel}}]{mayer2004}
{Mayer}, L., {Quinn}, T., {Wadsley}, J., \& {Stadel}, J. 2004, ApJ, 609, 1045

\bibitem[{{Mej{\'{\i}}a} {et~al.}(2005){Mej{\'{\i}}a}, {Durisen}, {Pickett}, \&
  {Cai}}]{mejia2005}
{Mej{\'{\i}}a}, A.~C., {Durisen}, R.~H., {Pickett}, M.~K., \& {Cai}, K. 2005,
  ApJ, 619, 1098

\bibitem[{{Monaghan}(1992)}]{monaghan1992}
{Monaghan}, J.~J. 1992, \araa, 30, 543

\bibitem[{{Nelson} {et~al.}(2000){Nelson}, {Benz}, \&
  {Ruzmaikina}}]{nelson_benz_ruzmaikina2000}
{Nelson}, A.~F., {Benz}, W., \& {Ruzmaikina}, T.~V. 2000, ApJ, 529, 357

\bibitem[{{Osterbrock}(1962)}]{osterbrock1962}
{Osterbrock}, D.~E. 1962, \apj, 136, 359

\bibitem[{{Pathria}(1996)}]{pathria}
{Pathria}, R.~K. 1996, Statistical Mechanics, Second Edition
  (Butterworh-Heinemann)

\bibitem[{{Pickett}(1995)}]{pickett1995}
{Pickett}, B.~K. 1995, Ph.D.~Thesis

\bibitem[{{Pickett} {et~al.}(1998){Pickett}, {Cassen}, {Durisen}, \&
  {Link}}]{pickett1998}
{Pickett}, B.~K., {Cassen}, P., {Durisen}, R.~H., \& {Link}, R. 1998, ApJ, 504,
  468

\bibitem[{{Pickett} {et~al.}(2000{\natexlab{a}}){Pickett}, {Cassen}, {Durisen},
  \& {Link}}]{pickett2000}
---. 2000{\natexlab{a}}, ApJ, 529, 1034

\bibitem[{{Pickett} {et~al.}(2000{\natexlab{b}}){Pickett}, {Durisen}, {Cassen},
  \& {Mejia}}]{pdcm2000}
{Pickett}, B.~K., {Durisen}, R.~H., {Cassen}, P., \& {Mejia}, A.~C.
  2000{\natexlab{b}}, ApJl, 540, L95

\bibitem[{{Pickett} {et~al.}(2003){Pickett}, {Mej{\'{\i}}a}, {Durisen},
  {Cassen}, {Berry}, \& {Link}}]{pickett2003}
{Pickett}, B.~K., {Mej{\'{\i}}a}, A.~C., {Durisen}, R.~H., {Cassen}, P.~M.,
  {Berry}, D.~K., \& {Link}, R.~P. 2003, ApJ, 590, 1060

\bibitem[{{Rafikov}(2005)}]{rafikov2005}
{Rafikov}, R.~R. 2005, ApJl, 621, L69

\bibitem[{{Rice} {et~al.}(2003){Rice}, {Armitage}, {Bate}, \&
  {Bonnell}}]{rice2003}
{Rice}, W.~K.~M., {Armitage}, P.~J., {Bate}, M.~R., \& {Bonnell}, I.~A. 2003,
  MNRAS, 339, 1025

\bibitem[{{Rice} {et~al.}(2005){Rice}, {Lodato}, \& {Armitage}}]{mnras364l56}
{Rice}, W.~K.~M., {Lodato}, G., \& {Armitage}, P.~J. 2005, MNRAS, 364, L56

\bibitem[{{Rice} {et~al.}(2004){Rice}, {Lodato}, {Pringle}, {Armitage}, \&
  {Bonnell}}]{rice2004}
{Rice}, W.~K.~M., {Lodato}, G., {Pringle}, J.~E., {Armitage}, P.~J., \&
  {Bonnell}, I.~A. 2004, MNRAS, 355, 543

\bibitem[{{Rodr{\'{\i}}guez-Fern{\'a}ndez}
  {et~al.}(2000){Rodr{\'{\i}}guez-Fern{\'a}ndez}, {Mart{\'{\i}}n-Pintado}, {de
  Vicente}, {Fuente}, {H{\"u}ttemeister}, {Wilson}, \&
  {Kunze}}]{rodriguez-fernandez2000}
{Rodr{\'{\i}}guez-Fern{\'a}ndez}, N.~J., {Mart{\'{\i}}n-Pintado}, J., {de
  Vicente}, P., {Fuente}, A., {H{\"u}ttemeister}, S., {Wilson}, T.~L., \&
  {Kunze}, D. 2000, \aap, 356, 695

\bibitem[{{Sano} {et~al.}(2000){Sano}, {Miyama}, {Umebayashi}, \&
  {Nakano}}]{sanoetal2000}
{Sano}, T., {Miyama}, S.~M., {Umebayashi}, T., \& {Nakano}, T. 2000, \apj, 543,
  486

\bibitem[{{Spitzer} \& {Tomasko}(1968)}]{spitzer_tomasko1968}
{Spitzer}, L.~J. \& {Tomasko}, M.~G. 1968, \apj, 152, 971

\bibitem[{{Stepinski}(1992)}]{stepinski1992}
{Stepinski}, T.~F. 1992, Icarus, 97, 130

\bibitem[{{Sternberg} \& {Neufeld}(1999)}]{sternberg_neufeld1999}
{Sternberg}, A. \& {Neufeld}, D.~A. 1999, \apj, 516, 371

\bibitem[{{Stone} \& {Norman}(1992)}]{stone_norman1992}
{Stone}, J.~M. \& {Norman}, M.~L. 1992, \apjs, 80, 753

\bibitem[{{Umebayashi} \& {Nakano}(1981)}]{umebayashi_nakano1981}
{Umebayashi}, T. \& {Nakano}, T. 1981, \pasj, 33, 617

\bibitem[{{Wadsley} {et~al.}(2004){Wadsley}, {Stadel}, \&
  {Quinn}}]{wadsleyetal2004}
{Wadsley}, J.~W., {Stadel}, J., \& {Quinn}, T. 2004, New Astronomy, 9, 137

\bibitem[{{Walmsley} {et~al.}(2004){Walmsley}, {Flower}, \& {Pineau des
  For{\^e}ts}}]{walmsleyetal2004}
{Walmsley}, C.~M., {Flower}, D.~R., \& {Pineau des For{\^e}ts}, G. 2004, \aap,
  418, 1035

\bibitem[{{Webber}(1998)}]{webber1998}
{Webber}, W.~R. 1998, \apj, 506, 329

\bibitem[{{Whitehouse} \& {Bate}(2006)}]{whitehouse2006}
{Whitehouse}, S.~C. \& {Bate}, M.~R. 2006, \mnras, 367, 32

\end{thebibliography}
\end{document}